\documentclass[pdflatex,sn-mathphys-num]{sn-jnl}

\usepackage{amsmath,amssymb,amsfonts}%
\usepackage{mathtools}
\usepackage{caption}
\usepackage{subcaption}
\usepackage{graphicx}%
\usepackage{multirow}%
\usepackage{amsthm}%
\usepackage{mathrsfs}%
\usepackage[title]{appendix}%
\usepackage{xcolor}%
\usepackage{textcomp}%
\usepackage{manyfoot}%
\usepackage{booktabs}%
\usepackage{algorithm}%
\usepackage{algorithmicx}%
\usepackage{algpseudocode}%
\usepackage{listings}%

\begin{document}

\title[Assessing Large Language Models for Automated Feedback Generation]{Assessing Large Language Models for Automated Feedback Generation in Learning Programming Problem Solving}

\author*[1,3]{\fnm{Priscylla} \sur{Silva}}\email{priscylla.silva@usp.br}

\author[2]{\fnm{Evandro} \sur{Costa}}\email{evandro@ic.ufal.br}

\affil*[1]{\orgname{University of São Paulo}, \orgaddress{\city{São Carlos}, \state{SP}, \country{Brazil}}}

\affil[2]{\orgname{Federal University of Alagoas}, \orgaddress{\city{Maceió}, \state{AL}, \country{Brazil}}}

\affil[3]{\orgname{Federal Institute of Alagoas}, \orgaddress{\city{Rio Largo}, \state{AL}, \country{Brazil}}}

\abstract{
Providing effective feedback is important for student learning in programming problem-solving. In this sense, Large Language Models (LLMs) have emerged as potential tools to automate feedback generation. However, their reliability and ability to identify reasoning errors in student code remain not well understood. This study evaluates the performance of four LLMs (GPT-4o, GPT-4o mini, GPT-4-Turbo, and Gemini-1.5-pro) on a benchmark dataset of 45 student solutions. We assessed the models' capacity to provide accurate and insightful feedback, particularly in identifying reasoning mistakes. Our analysis reveals that 63\% of feedback hints were accurate and complete, while 37\% contained mistakes, including incorrect line identification, flawed explanations, or hallucinated issues. These findings highlight the potential and limitations of LLMs in programming education and underscore the need for improvements to enhance reliability and minimize risks in educational applications.
}


\maketitle

\section{Introduction}
\label{sec:intro}

Despite its foundational role in computing education, computer programming remains a difficult subject for many learners. Programming courses often enroll large numbers of students, making it difficult for instructors to offer timely and personalized feedback. This feedback is crucial for helping learners understand complex concepts, correct misconceptions, and develop problem-solving skills. However, generating such feedback for programming assignments are particularly demanding. Instructors must evaluate whether a solution meets the problem's requirements, understand the student's thought process, identify errors, and communicate guidance to foster reflection and learning~\citep{silva-2019-its}.

Automated feedback systems have emerged as a solution to address the challenges of assessing student submissions. Traditional tools typically rely on predefined rules or rigid evaluation frameworks~\citep{deeva-2021}. While these methods can be effective in specific contexts, they lack the flexibility to provide individualized and context-aware feedback~\citep{maier-2022-100080}. Recent studies have demonstrated that large language models (LLMs) can be used to overcome this limitation by providing natural student feedback that adapts to each student's specific needs~\citep{kiesler-2023-fie}.

However, using LLMs in educational contexts raises significant concerns. \citet{tyen-etal-2024-llms} show that LLMs struggle with reasoning tasks, such as accurately following the step-by-step resolution of arithmetic expressions. These models show less than 50\% accuracy in both arithmetic and logical deduction tasks, particularly in identifying and correcting errors, as they often fail to indicate the specific step where a mistake occurred. Furthermore, large language models (LLMs) are susceptible to generating hallucinations or providing misleading information, which may frustrate students and interfere with the learning process~\citep{jukiewicz-2024-future}. Such inaccuracies are particularly concerning in education, where trust and reliability are paramount.

In this work, we address these challenges by providing a systematic evaluation of four different LLMs (GPT-4o, GPT-4o-mini, GPT-4-Turbo, and Gemini-1.5-pro) on their ability to generate accurate and meaningful feedback for programming assignments. Unlike prior studies that typically focus on a single LLM, Our comparative analysis highlights the different capabilities and limitations of multiple LLMs. To ensure reproducibility, we contribute a benchmark dataset based on real-world student submissions from introductory programming courses, making it publicly available to foster further research in this area.

In summary, the main contributions of this work are:
\begin{enumerate}
    \item A evaluation of four LLMs in generating feedback for programming assignments, focusing on their ability to identify and explain student mistakes;
    \item A publicly available benchmark dataset of annotated student code submissions, facilitating future research on automated feedback systems.
\end{enumerate}

\section{Related Work}

\textbf{Evaluating Feedback Generated by LLMs in Programming Tasks}. \citet{Pankiewicz2023} used GPT-3.5 to provide hints to students upon request during programming tasks when errors such as compilation failures, runtime issues, or unit test failures were identified. Their approach used prompts that included the problem description, student code, and error details. Feedback usefulness was evaluated based on student opinions using a Likert scale. However, the study did not assess whether the feedback correctly aligned with the problem requirements or addressed issues such as hallucinations or logical inconsistencies.

\citet{Azaiz-2024-feedback} examined GPT-4 Turbo for feedback generation on 55 solutions to two programming assignments. They reported feedback accuracy rates of 75\% to 95\% in determining the correctness of solutions but noted that 22\% of the feedback was inconsistent or incorrect. Similarly, \citet{Roest-2024-nextstephint} evaluated GPT-3.5 Turbo for real-time hint generation during students' problem-solving processes. Experts reviewed 48 hints and found that 33\% contained misleading information. \citet{Jacobs-2024-feedback} employed GPT-4 to generate feedback using task specifications, student code, compiler output, and unit test results. Expert evaluation of 137 feedbacks revealed that 12\% contained incorrect suggestions, and 6\% exhibited hallucinations.

These studies underscore the challenges in ensuring the accuracy and reliability of LLM-generated feedback. Although the proportion of issues in feedback (ranging from 6\% to 33\%) may appear small, in the context of education, even a small rate of misleading feedback can negatively affect students' learning, leading to misunderstandings, frustration, and loss of trust in the system. While these studies focus on evaluating individual models and their ability to generate feedback, our research goes further by conducting a systematic comparison of multiple LLMs.

\textbf{Challenges in Evaluating Reasoning with LLMs}.
Detecting reasoning mistakes in student code requires understanding the underlying logic and intent of the solution, going beyond surface-level syntax or test failures. Recent research has shown that LLMs generally perform poorly in identifying logical mistakes compared to their ability to correct them. \citet{tyen-etal-2024-llms} demonstrated that several LLMs struggle with logical deduction, arithmetic expressions, and word-ordering tasks, often failing to pinpoint where reasoning mistakes occur. Similarly, \citet{xia-2024-evaluating} observed similar limitations in mathematical reasoning tasks, where LLMs frequently misinterpret problem-solving logic.

\section{Benchmark} \label{sec:benchmark}

We created a benchmark dataset based on real-world student submissions to evaluate LLMs' performance in providing feedback for programming tasks. This dataset comprises 45 Python solutions to 5 introductory programming assignments collected from an online system used in programming courses. These 45 solutions were submitted by 5 students, with some students providing multiple solutions for the same programming assignment. This reflects a range of solution attempts and iterative improvements, capturing diverse solution strategies and common errors. Each assignment was designed to test foundational programming concepts such as loops, conditionals, and functions.

\textbf{Dataset Construction}.
The dataset was generated from student submissions to an automated grading system, which determined the correctness of each solution based on problem requirements and test cases. The students' solutions were written in Python 3.11. Table~\ref{tab:dataset} shows an overview of the dataset collected. This dataset was used to assess LLMs' feedback and evaluate their ability to identify mistakes and suggest improvements and corrections.

\begin{table}[h]
  \caption{Distribution of students' solutions across assignments and correctness based on the submission system.} \label{tab:dataset}
  \begin{tabular}{lccc}
  \toprule
  \bfseries Assignment & \bfseries Total Solutions & \bfseries Correct & \bfseries Incorrect\\
  \midrule
  Area of a Circle & 14 & 05  & 09 \\
  Simple Sum & 06 & 04  & 02 \\
  T-Shirts & 10 & 02  & 08 \\
  Huaauhahhuahau & 08 & 04  & 04 \\
  Grandpa's Balance & 07 & 04  & 03 \\
  \midrule
   \textbf{Total} & 45 & 19  & 26 \\
  \bottomrule
  \end{tabular}
\end{table}

\begin{figure}[ht]
\centering
\includegraphics[width=1\textwidth]{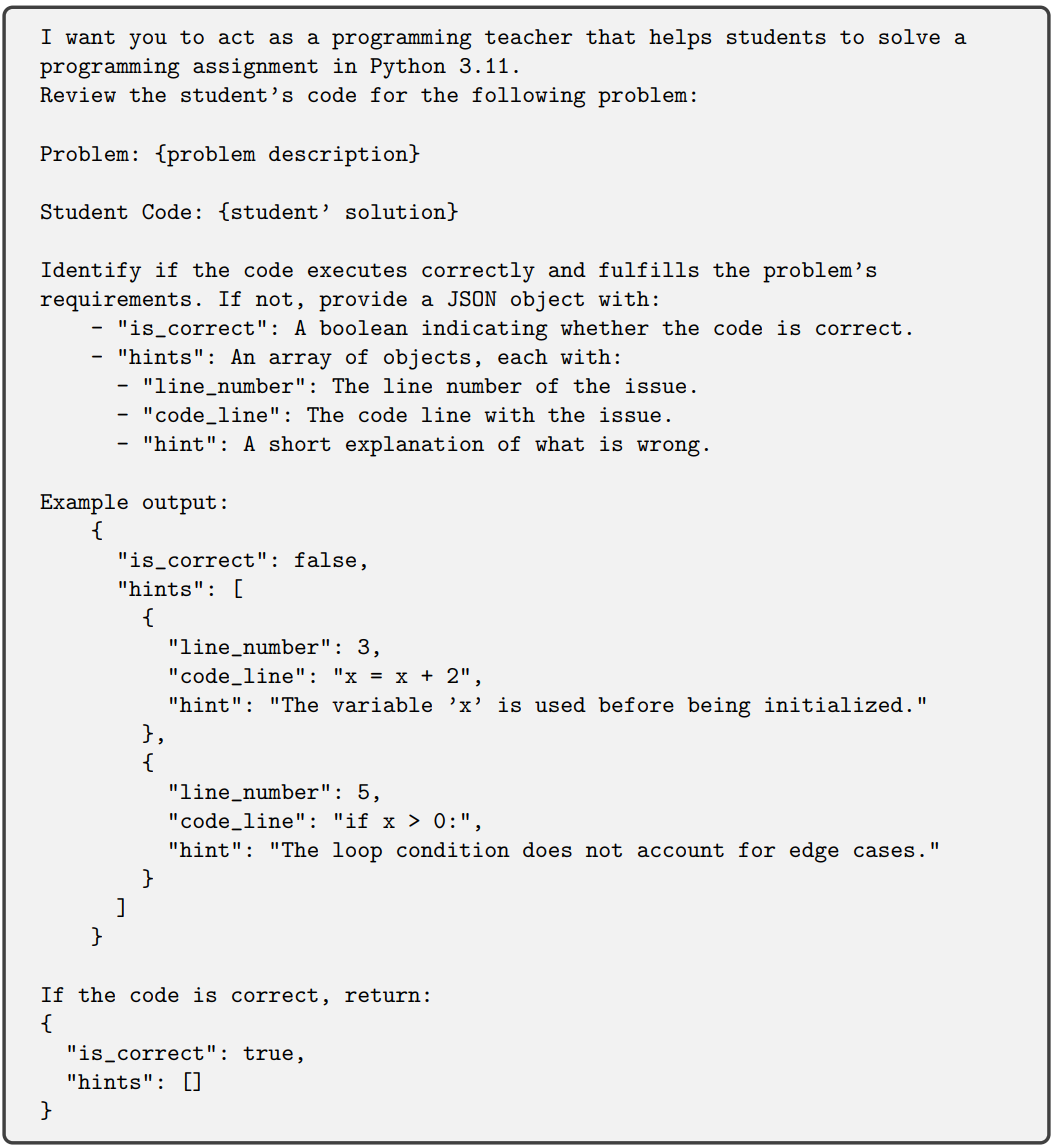}
\caption{Prompt used to ask for feedback.}\label{fig:promp}
\end{figure}

\textbf{Feedback Generation}. Feedback was generated using four LLMs: GPT-4o (version: 2024-08-06), GPT-4o-mini (version: 2024-07-18), GPT-4-Turbo (version: 2024-04-09), and Gemini-1.5-pro. Each model was prompted using the same template to ensure consistency in feedback generation (see the prompt template in Fig. \ref{fig:promp}).
The models were executed with a temperature setting of 0 (zero) to try to ensure consistent output results. The prompt explicitly instructed the models to act as programming teachers, assess the correctness of the code, and generate JSON-formatted feedback. Feedback included a binary indicator (\emph{is\_correct}) and a list of hints, with each hint specifying:

\begin{itemize}
    \item The line number of the issue,
    \item The code line containing the issue,
    \item A concise explanation of the problem.
\end{itemize}

\textbf{Feedback Annotation}. Two teaching assistants annotated the feedback generated by the models. The feedback was assessed in two stages:

\begin{enumerate}
    \item \textbf{Automatic Annotation}: The correctness of the solutions was determined using the results from the automated submission system.
    \item \textbf{Human Annotation}: Teaching assistants evaluated the feedback and organized it into five categories. Disagreement cases were resolved through discussion and consensus.
\end{enumerate}

Each hint was categorized individually, following an adaptation of the framework by \citet{hellas-2023-exploring}. In our categorization, any error in the hint message makes it incorrect, even if other parts are accurate. The categories are defined as follows:

\begin{itemize}

    \item \textbf{Accurate and Complete (AC)}: The feedback correctly identifies the appropriate line of code and explains the issue and/or how to resolve it. This is considered the ideal type of feedback, as it effectively supports students in identifying and understanding their mistakes.
    
    \item \textbf{Accurate Line Only (ALO)}: The feedback correctly identifies the line containing the issue but provides an incorrect or misleading explanation of the problem. Unlike False Positive (FP), the line identified does indeed contain a mistake, but the model misunderstands the nature of the issue.
    
    \item \textbf{Accurate Issue Only (AIO)}: The feedback describes the right issue but does not indicate the correct line of code. This misalignment can confuse students, as the explanation may not appear to relate to the line they are working on.
    
    \item \textbf{False Positive (FP)}: The feedback incorrectly identifies an issue where none exists. There is no actual error on the identified line, and the feedback message is entirely fake (hallucinated). This type of feedback can undermine student trust in the system.
    
    \item \textbf{Misleading Suggestions} (MS): The feedback points out a real error in the code, but both the line and the suggested fix are incorrect. While related to AIO, MS adds an additional layer of confusion by including a flawed or misleading resolution.    
\end{itemize}

Table \ref{tab:annotation} displays the categorization of the hints produced by each model, indicating the frequency of each category. The analysis focuses only on true negatives, which means that the dataset includes only the feedback for solutions that the models accurately identified as incorrect.

\begin{table}[hbtp]
  \caption{Categorization of hints generated by each model.} \label{tab:annotation}
  \begin{tabular}{lcccccc}
  \toprule
  \multirow{2}{4em}{\textbf{Model}}& \multicolumn{6}{c}{\textbf{Categories}} \\
   & \bfseries AC & \bfseries ALO & \bfseries AIO & \bfseries FP & \bfseries MS & \bfseries Total \\
  \midrule
  GPT-4o-mini & 36 & 02  & 06 & 03 & 02 & 49 \\
  GPT-4o & 28 & 10  & 03 & 08 & 00  & 49 \\
  GPT-4-Turbo & 33 & 06  & 02 & 03 & 00 & 44 \\
  Gemini-1.5-pro & 23 & 09  & 07 & 09 & 00 & 48 \\
  \midrule
   \textbf{Total} & 120 & 27  & 18 & 23 & 02 & 190 \\
  \bottomrule
  \end{tabular}
\end{table}

\section{Evaluation of Model Performance}

To assess the ability of LLMs to evaluate student code, we use our benchmark dataset to investigate two key research questions:
\begin{itemize}
    \item[\textbf{RQ1.}] How accurately do LLMs determine whether a student's solution fulfills the requirements of a programming task?

    \item[\textbf{RQ2.}] What are the common types of errors in student code that LLMs fail to identify?
\end{itemize}

\subsection{Results for RQ1: Accuracy in Correctness Evaluation}

\begin{figure}[htbp]
  \caption{Confusion matrices for GPT-4o, GPT-4o-mini, and GPT-4-Turbo, showing the models' performance in classifying correct and incorrect student solutions.} \label{fig:confusion_matrix}
  \includegraphics[width=1\linewidth]{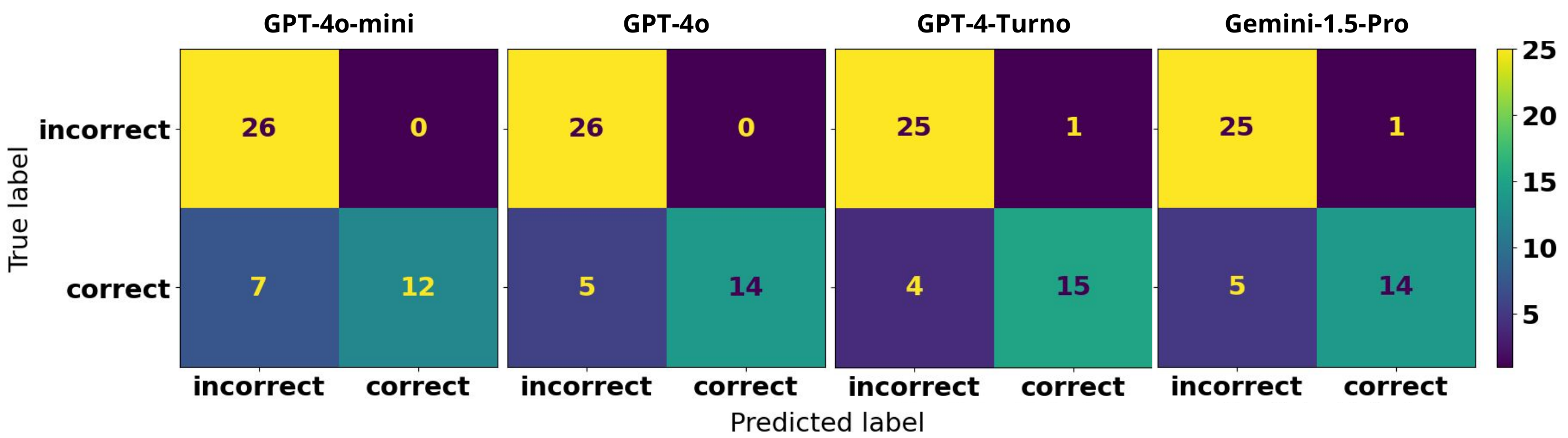}
\end{figure}

The models performed similarly in evaluating the correctness of students' code (Fig. \ref{fig:confusion_matrix} shows a visualization of the confusion matrices for each model). The accuracy metrics for each model are as follows: GPT-4o-mini: 84.44\%, Gemini-1.5-pro: 86.67\%, GPT-4o: 88.89\%, and GPT-4-Turbo: 88.89\%.

\textbf{GPT-4o-mini} misclassified seven correct solutions as incorrect. These errors were distributed across three tasks: four from the \textit{Simple Sum} assignment, two from the \textit{T-shirt} assignment, and one from \textit{Grandpa's Balance}. In the \textit{Simple Sum} task, the model incorrectly claimed the solutions did not meet the required output format, despite their correctness. Similarly, in the \textit{T-shirt} and \textit{Grandpa's Balance} assignments, the model misinterpreted the students' logical reasoning, falsely identifying errors in the code.

\textbf{GPT-4o} exhibited similar misclassification patterns, with five correct solutions identified as incorrect. These were a subset of the errors made by GPT-4o-mini, including three \textit{Simple Sum} solutions and two \textit{T-shirt} solutions. GPT-4o successfully identified that one solution from \textit{Simple Sum} met the output requirements, which GPT-4o-mini failed to recognize. However, the model also struggled to interpret logical reasoning in the \textit{T-shirt} task.

\textbf{GPT-4-Turbo} demonstrated comparable accuracy to GPT-4o, misclassifying four correct solutions as incorrect. It improved over GPT-4o-mini in the \textit{Simple Sum} assignment, reducing errors to two cases. However, like the others, it struggled with the logical reasoning in two \textit{T-shirt} assignment solutions.

\textbf{Gemini-1.5-pro} exhibited a similar pattern of errors as the GPT models. The misclassified solutions were a subset of those incorrectly labeled by GPT-4o and GPT-4o-mini. Gemini and GPT-4-Turbo were the only ones to classify an incorrect solution as correct incorrectly.

\subsection{Results for RQ2: Error Detection and Feedback Quality}

\begin{figure}[h]
    \centering
  \caption{Comparison of the frequency of feedback categories generated by each model.} \label{fig:barchart}
  \includegraphics[width=0.7\linewidth]{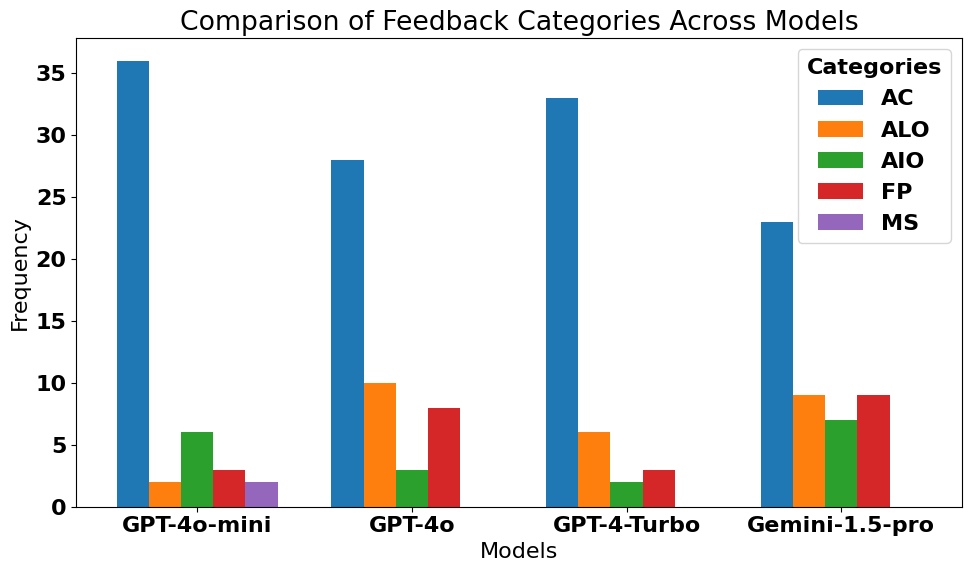}
\end{figure}

To evaluate the quality of feedback generated by the models, we categorized hints associated with correctly identified incorrect solutions into five categories: Accurate and Complete (AC), Accurate Line Only (ALO), Accurate Issue Only (AIO), False Positive (FP), and Misleading Suggestions (MS). Fig. \ref{fig:barchart} illustrates the distribution of feedback categories across models.

GPT-4o-mini generated the highest number of \textit{Accurate and Complete} hints, followed closely by GPT-4-Turbo. These results indicate the strength of these models in providing both the correct line and the proper explanation of the mistake. Furthermore, GPT-4o and Gemini-1.5-pro generated more \textit{Accurate Line Only} hints, suggesting a tendency to identify the line containing the issue but not adequately explaining the problem or assisting the student in fixing it. Across all models, the frequency of \textit{False Positives} was relatively low, occurring in approximately 12\% of cases. These errors were more frequent in Gemini-1.5-pro and GPT-4o. Similarly, \textit{Misleading Suggestions} were rare, with only two instances reported for GPT-4o-mini.

\section{Limitations and Future Directions}

This study has potential limitations. First, the benchmark dataset comprises 45 solutions from 5 students across five programming assignments. Although it captures diverse solution strategies and common errors, the dataset's small size and limited participant pool restrict the generalizability of our findings. Furthermore, imbalances in the number of submissions across assignments may introduce bias in the evaluation results. We plan to expand the dataset with more participants, diverse assignments, and augmented data to address these issues.

Second, the prompt design may have impacted the models' performance. In future studies, we will explore alternative prompt designs, such as separating tasks into individual prompts or using dynamic, task-specific prompts. This could provide a more accurate evaluation of LLM capabilities.

Finally, our findings are based on the evaluation of four specific LLMs. While these models represent state-of-the-art capabilities, the results may not generalize to other models or future iterations. Additionally, while human annotations were validated to ensure consistency, they may still introduce an element of subjectivity.

\section{Discussion and Conclusion}

In this work, we evaluate the ability of large language models (LLMs) to provide feedback on programming tasks. We introduce a benchmark dataset of student submissions in introductory programming courses and use it to analyze and compare the performance of GPT-4o, GPT-4o-mini, GPT-4-Turbo, and Gemini-1.5-pro. Our study categorizes feedback generated by these models into five categories, highlighting their strengths and limitations in providing proper feedback. The results show that while these LLMs are effective in generating accurate feedback in many cases, they also exhibit significant challenges, including hallucinated errors, misleading suggestions, and struggles to interpret student logic.

Overall, 63\% of the feedback hints generated by the models were totally correct, meaning they accurately identified the problematic line and provided an appropriate explanation. However, 37\% of the feedback contained some issues, such as pointing to the wrong line, providing an incorrect hint message, or hallucinating non-existent errors.

Our findings reveal that although LLMs perform well in detecting syntactic and surface-level issues, they often fail to capture deeper reasoning mistakes in student code. Additionally, the frequency of hallucinations and false positives highlights the need for further refinement of these models to minimize their risks in educational settings. The benchmark, prompts, and supplementary materials can be found at: \url{https://github.com/priscylla/Assessing-LMMs-for-Feedback-Generation}.


\bibliography{main}

\end{document}